\PassOptionsToPackage{unicode}{hyperref}
\PassOptionsToPackage{hyphens}{url}
\documentclass[
]{article}
\usepackage{xcolor}
\usepackage[margin=3cm]{geometry}
\usepackage{amsmath,amssymb}
\usepackage{iftex}
\ifPDFTeX
  \usepackage[T1]{fontenc}
  \usepackage[utf8]{inputenc}
  \usepackage{textcomp} % provide euro and other symbols
\else % if luatex or xetex
  \usepackage{unicode-math} % this also loads fontspec
  \defaultfontfeatures{Scale=MatchLowercase}
  \defaultfontfeatures[\rmfamily]{Ligatures=TeX,Scale=1}
\fi
\usepackage{lmodern}
\ifPDFTeX\else
\fi
\IfFileExists{upquote.sty}{\usepackage{upquote}}{}
\IfFileExists{microtype.sty}{% use microtype if available
  \usepackage[]{microtype}
  \UseMicrotypeSet[protrusion]{basicmath} % disable protrusion for tt fonts
}{}
\makeatletter
\@ifundefined{KOMAClassName}{% if non-KOMA class
  \IfFileExists{parskip.sty}{%
    \usepackage{parskip}
  }{% else
    \setlength{\parindent}{0pt}
    \setlength{\parskip}{6pt plus 2pt minus 1pt}}
}{% if KOMA class
  \KOMAoptions{parskip=half}}
\makeatother
\usepackage{graphicx}
\makeatletter
\newsavebox\pandoc@box
\newcommand*\pandocbounded[1]{% scales image to fit in text height/width
  \sbox\pandoc@box{#1}%
  \Gscale@div\@tempa{\textheight}{\dimexpr\ht\pandoc@box+\dp\pandoc@box\relax}%
  \Gscale@div\@tempb{\linewidth}{\wd\pandoc@box}%
  \ifdim\@tempb\p@<\@tempa\p@\let\@tempa\@tempb\fi% select the smaller of both
  \ifdim\@tempa\p@<\p@\scalebox{\@tempa}{\usebox\pandoc@box}%
  \else\usebox{\pandoc@box}%
  \fi%
}
\def\fps@figure{htbp}
\makeatother
\NewDocumentCommand\citeproctext{}{}

\makeatletter
 \let\@cite@ofmt\@firstofone
 \def\@biblabel#1{}
 \def\@cite#1#2{{#1\if@tempswa , #2\fi}}
\makeatother
\newlength{\cslhangindent}
\newlength{\csllabelwidth}
\newenvironment{CSLReferences}[2] % #1 hanging-indent, #2 entry-spacing
 {\begin{list}{}{%
  \setlength{\itemindent}{0pt}
  \setlength{\leftmargin}{0pt}
  \setlength{\parsep}{0pt}
  \ifodd #1
   \setlength{\leftmargin}{\cslhangindent}
   \setlength{\itemindent}{-1\cslhangindent}
  \fi
  \setlength{\itemsep}{#2\baselineskip}}}
 {\end{list}}
\usepackage{calc}

\usepackage{authblk}
\usepackage{orcidlink}

\usepackage{bookmark}
\IfFileExists{xurl.sty}{\usepackage{xurl}}{} % add URL line breaks if available
\hypersetup{
  pdftitle={Blini: lightweight nucleotide sequence search and dereplication},
  hidelinks,
  pdfcreator={LaTeX via pandoc}}

\title{Blini: lightweight nucleotide sequence search and dereplication}
\author[1]{Amit Lavon\orcidlink{0000-0003-3928-5907}}
\newcounter{iaffil}
\affil[\theiaffil]{University of California, Irvine, CA, USA}
\stepcounter{iaffil}
\date{24 June 2025}

\begin{document}
\maketitle
\begin{abstract}
Blini is a tool for quick lookup of nucleotide sequences in databases,
and for quick dereplication of sequence collections. It is meant to help
clean and characterize large collections of assembled contigs or long
sequences that would otherwise be too big to search with online tools,
or too demanding for a local machine to process. Benchmarks on simulated
data demonstrate that it is faster than existing tools and requires less
RAM, while preserving search and clustering accuracy.
\end{abstract}

\section{Introduction}\label{introduction}

Metagenomes are collections of genetic material from various organisms,
which are often not initially known. Characterizing the taxonomic makeup
of a sample involves searching its contents in large databases in order
to find which organism matches each nucleotide sequence. Assembled
sequences can reach lengths of millions of bases, making alignment-based
search methods too cumbersome. Such big queries are often outsourced to
powerful cloud-based services such as BLAST (Altschul et al. 1990). In
recent years, k-mer-based algorithms were introduced that enabled
efficient searching in large datasets on local machines. Mash distance
(Ondov et al. 2016) introduced an alignment-free estimation formula for
average nucleotide identity between sequences, making sequence
comparison linear. Sourmash (Brown and Irber 2016) uses fractional
min-hashing in order to create small representations of large sequences,
which allow for efficient searching and comparison. The LinClust
clustering algorithm (Steinegger and Söding 2018) uses k-mer matching to
reduce the number of pairwise comparisons and achieve linear scaling
with the size of the input.

While each of these techniques provides value on its own, processing
datasets of tens or hundreds of thousands of genomes can still take many
hours on a local machine. Here, insights from previous techniques are
combined to create a unified tool that can search and dereplicate big
datasets quickly using estimated identity or containment.

\section{Results}\label{results}

\subsection{Algorithm}\label{algorithm}

\subsubsection{Fingerprinting with
k-mers}\label{fingerprinting-with-k-mers}

Blini uses constant-length subsequences (k-mers) to create fingerprints
for sequences. It uses the fractional min-hashing technique, similarly
to Sourmash (Brown and Irber 2016). A sliding window of length \(k\)
goes over the sequence and hashes each canonical \(k\)-long subsequence.
This collection of hashes is often called the sequence's \emph{sketch}.
The lower \(1/s\) hashes are retained, for an input scale parameter
\(s\). A high \(s\) means fewer hashes used in downstream calculations,
trading accuracy for better CPU and RAM performance. Once k-mer hashes
are extracted, the sequence is discarded and only its sketch is used for
downstream operations. These sketches can be saved to files and reused.

\subsubsection{Similarity estimation}\label{similarity-estimation}

Blini uses Mash distance (Ondov et al. 2016) to estimate average
nucleotide identity (ANI) between sequences. This formula translates the
Jaccard similarity between two k-mer sets to an estimation of the ANI
between the original sequences. For containment matching, the hashes of
the query sequence are compared against their intersection with the
hashes of the reference sequence.

\subsubsection{Search}\label{search}

The first step of searching is indexing the reference dataset. After the
reference sequences are fingerprinted, the 25\% lowest hashes are used
for indexing. The index is a mapping from hash value to a list of
sequence identifiers of the reference sequences that had that hash in
their fingerprints. The number 25\% was chosen as a sweet spot between
saving memory and retaining enough information for the search. As an
optimization, hashes with a single reference sequence are kept in a
separate `singletons' map. Since in practice most of the index elements
are singletons, this helps reduce RAM consumption and garbage collection
times.

In the second stage, each query sequence is read and fingerprinted. The
hash values are looked up in the index, and candidate reference
sequences are fetched. Then, the query sequence is compared against each
candidate sequence using Mash distance, and matches that pass the
similarity threshold are reported.

\subsubsection{Clustering}\label{clustering}

The clustering (dereplication) procedure follows a similar scheme to
LinClust (Steinegger and Söding 2018). Sequences are indexed and ordered
from the longest to the shortest. Then, going by that order, each
sequence is searched for using the search procedure. Matches that pass
the similarity threshold are joined with the query sequence and are
considered a cluster. These matches are then removed from the search
loop's candidates. This clustering procedure does not produce
inter-cluster distances for hierarchy generation.

\subsection{Performance evaluation}\label{performance-evaluation}

\subsubsection{Search - small}\label{search---small}

The search function was tested on RefSeq's viral reference (Pruitt,
Tatusova, and Maglott 2007). Blini was compared against Sourmash and
MMseqs. 100 viral genomes were randomly selected for the test. The
algorithms were then run on the 100 genomes as queries, and the original
database as reference. Each algorithm was expected to match each genome
to its source in the database. In a second run, random SNPs were
introduced to 1\% of the genomes' bases, and the same test was rerun.
For each test, the number of matches with sequences other than the
query's source was also measured. The searches were run against an index
of the reference dataset, created by each tool.

All three tools were able to match all 100 queries with their sources in
the database (Figure 1a). The number of non-source matches was 824 and
712 in Blini, 865 and 660 in Sourmash, and 3143 and 3019 in MMseqs, in
the raw and mutated datasets, respectively (Figure 1b).

Run-time was measured for searching the 100 sequences sequentially.
Blini and MMseqs were executed once and searched for all the queries in
one run, while Sourmash had to be executed once for each individual
query. Each run was repeated five times and the average run-time is
reported. Blini completed the run in 0.5 seconds, Sourmash completed the
run in 126 seconds, and MMseqs completed the run in 151 seconds (Figure
1c). The times shown here do not include reference-preprocessing time.

\begin{figure}
\centering
\pandocbounded{\includegraphics[keepaspectratio]{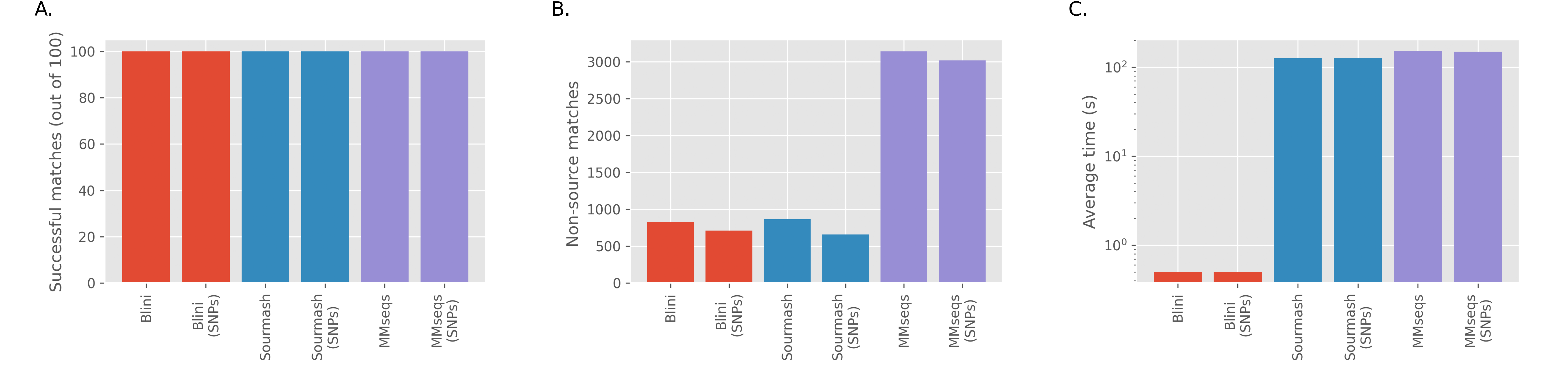}}
\caption{Search results for the viral dataset. Each tool was run on 100
randomly chosen viral genomes, to find them in the original dataset. (A)
shows how many of the 100 genomes were correctly mapped to their source
in the database. (B) shows how many additional matches were found in
addition to the 100 chosen genomes. (C) shows the search times for the
100 queries together.}
\end{figure}

\subsubsection{Search - big}\label{search---big}

To test the search function on a large dataset, the bacterial contigs
from (Pasolli et al. 2019) were used. This 10GB dataset contains 934K
contigs from almost 5K bacterial species. Each of the compared tools was
run to create an index of the dataset.

The simulated query dataset consisted of 100K random fragments of length
10K bases, sampled uniformly from the bacterial contigs. Each fragment
was mutated with random SNPs in 0.1\% of its bases. Blini, Sourmash and
MMseqs were run on the query dataset, to search it in the bacterial
reference. Because of the long search times, only Blini was run on the
full set of queries, while the other tools were run on one or ten
queries out of the 100K.

MMseqs took longer than 30 minutes to search for a single query, and was
therefore terminated prematurely. Sourmash was run on one query and on
ten queries and took 31 seconds per query. Blini took 6 seconds for one
and ten queries, and 25 seconds for the entire set of 100K queries
(Figure 2). This means a throughput of 5100 queries per second after the
6 seconds of loading the reference index. Blini matched all 100K queries
with their correct source in the reference, with 2444 additional
non-source matches (false-positives).

\begin{figure}
\centering
\includegraphics[width=0.4\linewidth,height=\textheight,keepaspectratio]{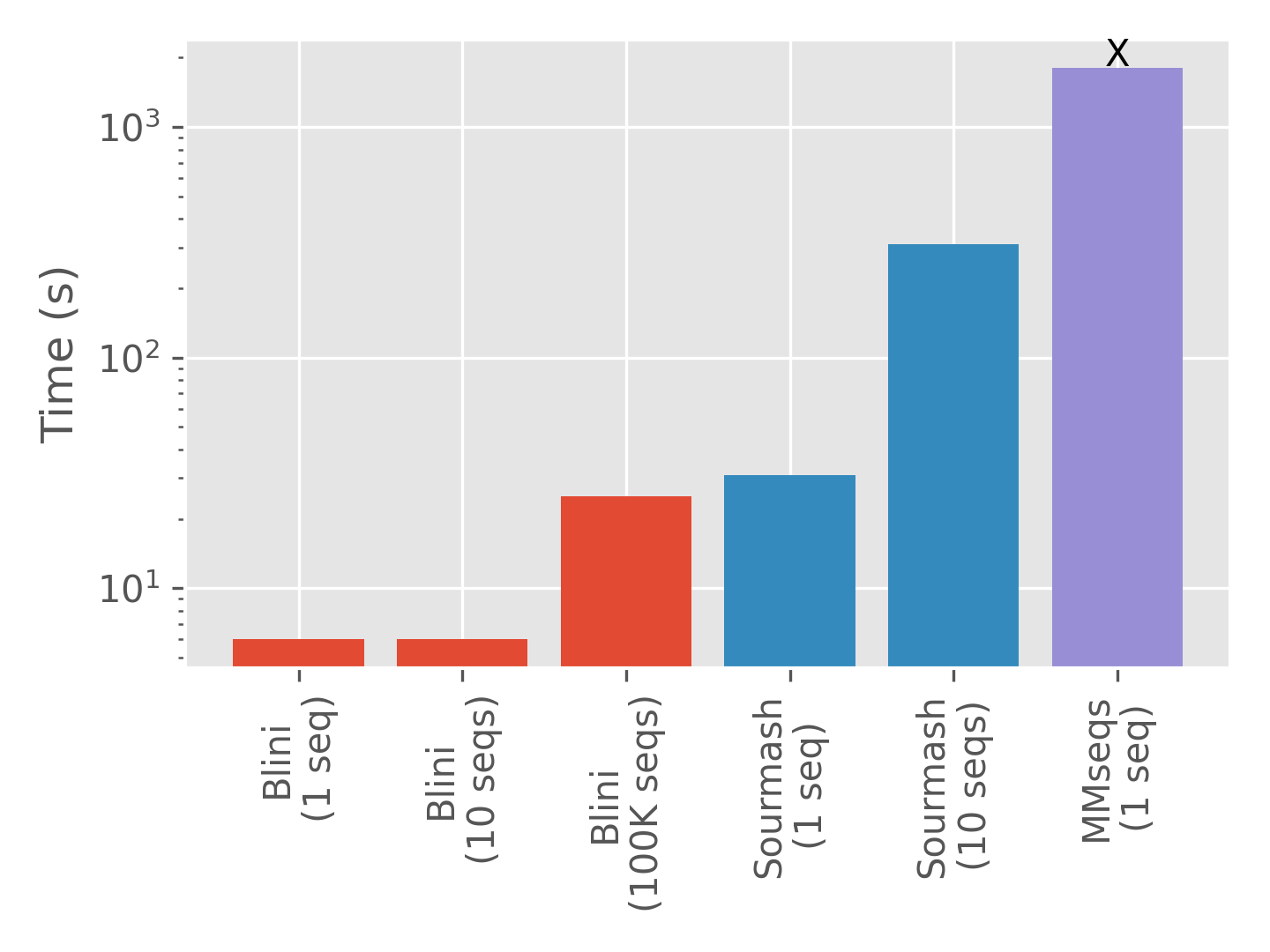}
\caption{Search times for the bacterial dataset. Each tool was run on
randomly chosen 10 kilo-base fragments from a 10GB bacterial dataset.
MMseqs is marked with an X because it was stopped manually before it
could finish running.}
\end{figure}

\subsubsection{Clustering}\label{clustering-1}

The clustering function was tested on two simulated datasets created
from the 100 chosen genomes of the previous test. In one dataset each
sequence had multiple counterparts with random SNPs. In the second
dataset random fragments were extracted from each root sequence. In the
SNPs dataset, each of the 100 original sequences had another 100 mutated
counterparts. Each counterpart had random SNPs in 1\% of its bases. In
the fragments dataset, each of the 100 original sequences had 300 random
fragments extracted from it, of length of at least 1000 bases. The
algorithms were expected to group each sequence with its mutated
counterparts or with its fragments. Performance was evaluated using the
Adjusted Rand-Index (ARI). Blini's \emph{scale} refers to the fraction
of k-mers considered for the operation. Scale 50 means that 1/50 of
k-mers were used.

In the SNPs dataset, both Blini and MMseqs achieved an ARI between 0.999
and 1.0, except for Blini with scale 200 which achieved an ARI of 0.997
(Figure 3b). Blini created 100, 100, 101 and 110 clusters using scales
25, 50, 100 and 200 respectively. MMseqs created 103 clusters (Figure
3a). Blini took on average 10.5 seconds, and MMseqs took 46 seconds with
one thread, and 14 seconds with four threads (Figure 3b). In terms of
memory, Blini had a maximal memory footprint of 255, 129, 65, and 38 MB
using scales 25, 50, 100 and 200 respectively. MMseqs had a maximal
memory footprint of 3072 MB (Figure 3c).

\begin{figure}
\centering
\includegraphics[width=0.8\linewidth,height=\textheight,keepaspectratio]{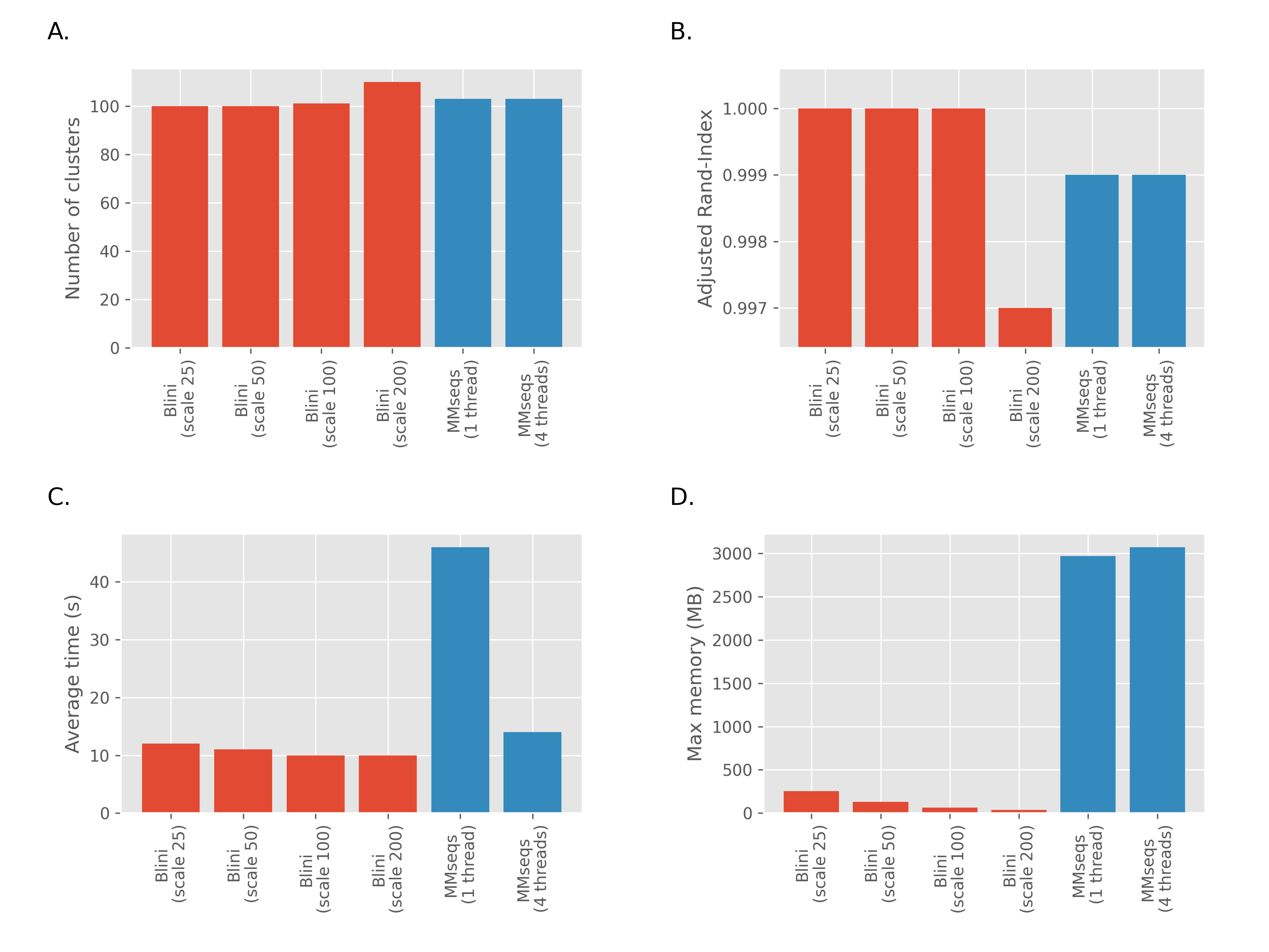}
\caption{Clustering results for the SNPs dataset. Each of the 100 viral
genomes from the search benchmark was used to create 100 mutant
sequences with SNPs in 1\% of their bases. The tools were run on this
collection of 10100 genomes and were expected to cluster them into 100
groups, corresponding to the original genomes.}
\end{figure}

In the fragments dataset, MMseqs achieved an ARI of 1.0 while Blini
achieved an ARI of 0.999, 0.999, 0.998 and 0.989 with scales 25, 50, 100
and 200 (Figure 4b). Blini grouped the dataset into 100, 104, 135 and
386 clusters, while MMseqs grouped the dataset into 101 clusters (Figure
4a). Blini took on average 20 seconds, and MMseqs took 80 seconds with
one thread, and 24 seconds with four threads (Figure 4c). In terms of
memory, Blini had a maximal memory footprint of 462, 233, 119, and 67 MB
using scales 25, 50, 100 and 200 respectively. MMseqs had a maximal
memory footprint of 5632 MB (Figure 4d).

\begin{figure}
\centering
\includegraphics[width=0.8\linewidth,height=\textheight,keepaspectratio]{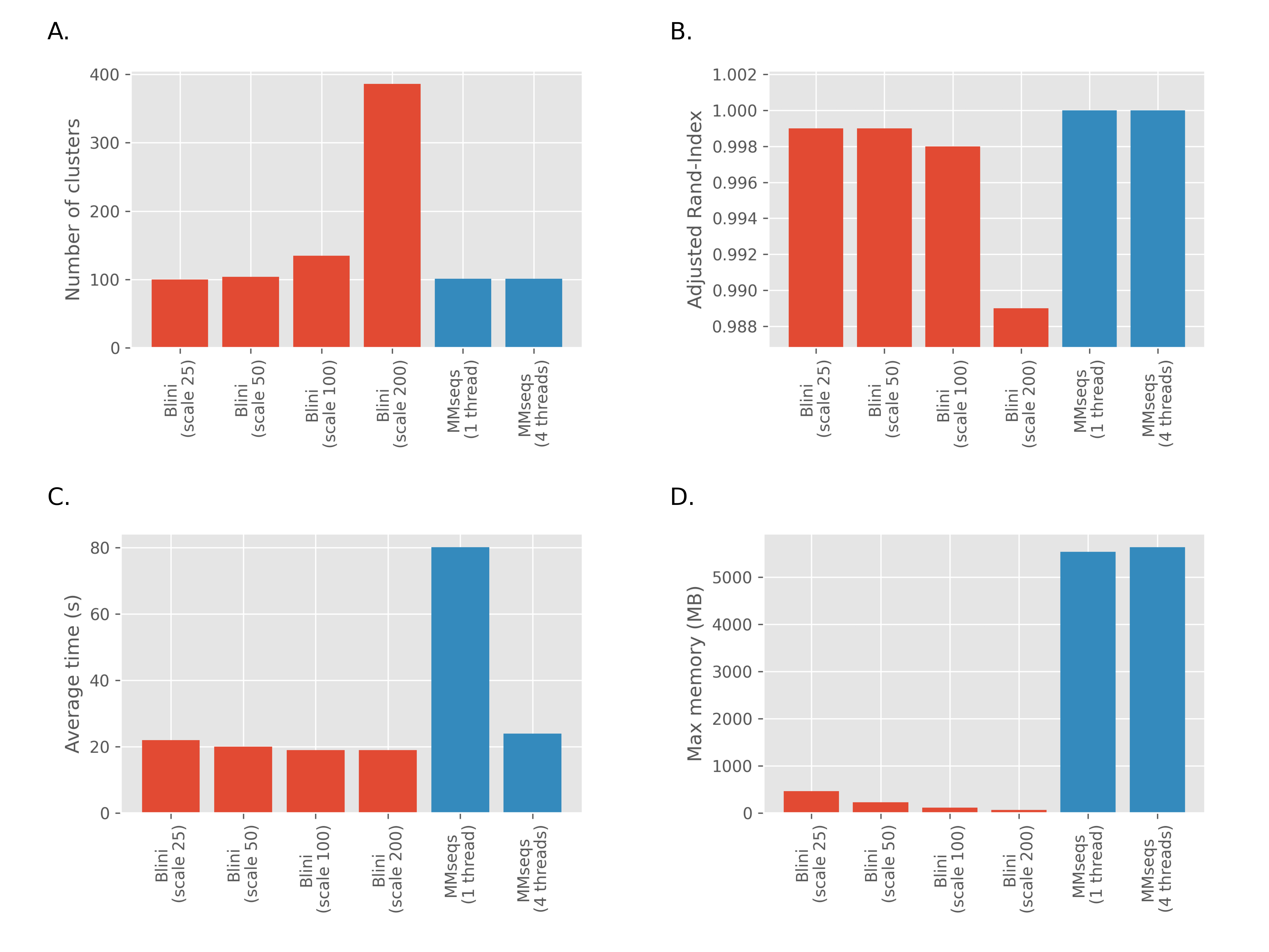}
\caption{Clustering results for the fragments dataset. Each of the 100
viral genomes from the search benchmark was used to create 300 random
fragments of length 1000 bases and above. The tools were run on this
collection of 30100 genomes and were expected to cluster them into 100
groups, corresponding to the original genomes.}
\end{figure}

\section{Discussion}\label{discussion}

This work introduces a new technique for nucleotide sequence search and
dereplication. The algorithm incorporates insights from several
developments in the field, specifically around the use of k-mers for
sequence search and average nucleotide identity (ANI) approximation,
combined with a minimalist implementation, resulting in a scalable and
easy to use utility.

The performance benchmarks shown here demonstrate an over 5000
queries-per-second throughput on a 10GB reference dataset, while
existing implementations took 30 seconds or over 30 minutes for a single
query. For clustering, the speedup was 4x in comparison to existing
implementations, as well as a reduced memory footprint, while clustering
accuracy was maintained.

One drawback is that Blini does not perform alignment, and instead uses
an approximation for ANI. This can be observed in Figure 3a-b, where
clustering accuracy decreased as the \emph{scale} parameter increased
beyond the recommended value for that dataset (see below). This approach
trades some accuracy for scalability.

The \emph{scale} parameter controls the ratio of k-mers used in the
procedure, and it can be tweaked. While higher scale values reduce CPU
and RAM consumption, they also affect the minimal length of sequences
the algorithm would be effective for. Blini is designed to work on
sequences \textasciitilde20 times longer than the selected \emph{scale}
value. For the default scale of 100, sequences shorter than 2000 are
likely to be falsely missed. This can be seen in Figure 3a-b, where a
decrease in clustering accuracy was observed for scale \textgreater50
with fragments of size 1000. While the scale can be tweaked, this tool
might not be suitable for short reads.

In summary, this is a new algorithm for search and dereplication of
large-scale sequence collections. Blini significantly reduces the
barrier to search and clustering tasks that would otherwise require huge
computing power to perform, while maintaining result accuracy. For
resource-constrained settings, Blini enables search and clustering that
would otherwise be impossible due to time and memory constraints. For
large systems such as cloud-based search servers, Blini can reduce query
costs significantly and open the door for increased bandwidth. The tool
is optimized for ease of use and does not require installing additional
software. Pre-sketched databases can be easily created, reused and
shared between research groups.

\section{Software Availability}\label{software-availability}

Code and runnable binaries are available for Linux, Windows and MacOS
at: https://github.com/fluhus/blini/releases

Feedback on the tool or on this manuscript is welcome at:
https://github.com/fluhus/blini/discussions

\section*{References}\label{references}
\addcontentsline{toc}{section}{References}

\protect\phantomsection\label{refs}


\begin{CSLReferences}{1}{0}
\bibitem[\citeproctext]{ref-altschul1990basic}
Altschul, Stephen F, Warren Gish, Webb Miller, Eugene W Myers, and David
J Lipman. 1990. {``Basic Local Alignment Search Tool.''} \emph{Journal
of Molecular Biology} 215 (3): 403--10.

\bibitem[\citeproctext]{ref-brown2016sourmash}
Brown, C Titus, and Luiz Irber. 2016. {``Sourmash: A Library for MinHash
Sketching of DNA.''} \emph{Journal of Open Source Software} 1 (5): 27.

\bibitem[\citeproctext]{ref-ondov2016mash}
Ondov, Brian D, Todd J Treangen, Páll Melsted, Adam B Mallonee, Nicholas
H Bergman, Sergey Koren, and Adam M Phillippy. 2016. {``Mash: Fast
Genome and Metagenome Distance Estimation Using MinHash.''} \emph{Genome
Biology} 17: 1--14.

\bibitem[\citeproctext]{ref-pasolli2019extensive}
Pasolli, Edoardo, Francesco Asnicar, Serena Manara, Moreno Zolfo,
Nicolai Karcher, Federica Armanini, Francesco Beghini, et al. 2019.
{``Extensive Unexplored Human Microbiome Diversity Revealed by over
150,000 Genomes from Metagenomes Spanning Age, Geography, and
Lifestyle.''} \emph{Cell} 176 (3): 649--62.

\bibitem[\citeproctext]{ref-pruitt2007ncbi}
Pruitt, Kim D, Tatiana Tatusova, and Donna R Maglott. 2007. {``NCBI
Reference Sequences (RefSeq): A Curated Non-Redundant Sequence Database
of Genomes, Transcripts and Proteins.''} \emph{Nucleic Acids Research}
35 (suppl\_1): D61--65.

\bibitem[\citeproctext]{ref-steinegger2018clustering}
Steinegger, Martin, and Johannes Söding. 2018. {``Clustering Huge
Protein Sequence Sets in Linear Time.''} \emph{Nature Communications} 9
(1): 2542.

\end{CSLReferences}
\end{document}